\documentclass[multphys,vecarrow]{svmult}

\usepackage{graphicx}
\usepackage{amssymb}
\usepackage{amsmath}
\usepackage{float}
\usepackage{subeqnar}
\usepackage{epsfig}
\usepackage{citesort}
\begin{document}
\frontmatter%%%%%%%%%%%%%%%%%%%%%%%%%%%%%%%%%%%%%%%%%%%%%%%%%%%%%%
%\input{BOOK/titlepage}
%\input{BOOK/preface}
%\tableofcontents
\mainmatter%%%%%%%%%%%%%%%%%%%%%%%%%%%%%%%%%%%%%%%%%%%%%%%%%%%%%%%

\bibliographystyle{unsrt}

\title*{Pedestrian Dynamics With Event-driven Simulation}
\author{Mohcine Chraibi\inst{1} and Armin Seyfried\inst{2}}
\institute{
Technical University Hamburg-Harburg\\
D-21071, Germany\\
Contact address, e-mail: {\tt mohcine.chraibi@tu-harburg.de}\\
\and
Central Institute for Applied Mathematics, Research Centre J\"ulich\\
52425 J\"ulich, Germany 
\\
Contact address, e-mail: {\tt a.seyfried@fz-juelich.de}\\
}

\date{\today}

\maketitle

\begin{abstract}
The social-force model is systematically modified to achieve a satisfying agreement with the fundamental diagram.
Furthermore, our modification allows an efficient computation of the simulation. 
Finally, different simulation-results will be compared to empirical data.
\\
\noindent
\end{abstract}

\section{Introduction}
Microscopic models are state of the art for computer simulation of pedestrian 
dynamics. The modeling of the individual movement of pedestrians is used to design a description of macroscopic pedestrian flow which allows e.g. the evaluation of 
escape routes, the design of pedestrian facilities and the study of more theoretical 
questions. For a first overview see \cite{Schreckenberg2001,Galea2003,Waldau2006}. 

One primary test, whether a model is appropriate for a quantitative description of pedestrian flow, is the comparison with the fundamental diagram \cite{Meyer-Konig2002,Kirchner2004,Hoogendoorn2001,RIMEA}.
\section{Social-force models}
 \subsection{General considerations}
The social-force model was introduced in \cite{Helbing1995}. It is expressed by the equation of motion 
\begin{equation}
m_i \frac{d^2 x_i}{d t^2} = m_i \frac{d v_i}{d t} = F_i = F^{drv}_i + F^{rep}_i + fluctuations.  
\label{forces}
\end{equation} 

where $x_i$ is the position of the $i^{th}$ pedestrian, $v_i$ its velocity and $m_i$ its mass.
\begin{equation}
F^{rep}_i =  \sum\limits_{j\neq i}^n F^{rep}_{ij}(x_j,x_i,v_i), n \in \mathbb N
\label{frep}
\end{equation}
  is the force due to interaction with other pedestrians. 
In the original model the introduction of the repulsive force $F^{rep}_i$ between pedestrians is motivated by the observation that pedestrians try to keep distance to others to secure their 'private sphere'. This behavior is also observed for the environment, i.e. pedestrians do not walk too close to walls and stairs. $F^{drv}_i$ is a driving term that makes the pedestrians attempt to move with their intended velocity. 

The term $fluctuations$ is added to take into account random variations of the behavior of pedestrians.  Additionally $fluctuations$ arise from deviations from the usual rules of motion. This term becomes important in high density situations \cite{Helbing1995}. 

\subsection{Simplified one-dimentional realization} 
For the modeling of pedestrian dynamics the pedestrians are represented by one-dimensional 'circles' with velocity dependent length $d_i$ and position (of the center of mass) $x_i$ moving in a continuous space.

An important aspect is the dependency between the current 
velocity and the required space of pedestrians \cite{SEYF05}. 
As mentioned in \cite{Seyfried2005} it is possible, on basis of empirical results, to determine the required length $d_i$ for the $i^{th}$ pedestrian, that moves in a single file with a velocity $v_i$, with $0.1\,m/s < v_i < 1.0\,m/s$: 
\begin{equation}
 d_i =  a + b\,v_i \quad \mbox{with} \quad a=0.36\,m \quad \mbox{and} \quad b=1.06\,s
 \label{dab}
\end{equation} 

One fundamental quantity in the model is the distance between the $i^{th}$ pedestrian and  the $j^{th}$ pedestrian which is defined as: 
\begin{equation}
dist_{i,j}(t)=\Delta (x_{i,j})-\frac{d_i+d_j}{2} 
\label{dist}
\end{equation}
with  $\Delta (x_{i,j}) = |x_i -  x_{j}|$ the distance between the center of mass of the $i^{th}$ pedestrian and the $j^{th}$ pedestrian.  $| . |$ denotes the absolute value in $\mathbb R$.  

\begin{figure}[ht]
\centerline{\epsfig{file=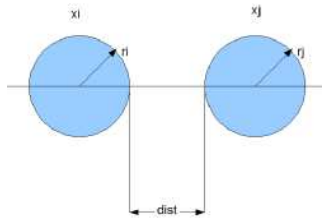,height=3cm,angle=0}}
\caption{Illustration of two pedestrians}
\label{fig:ccc}
\end{figure}

The velocity $v_i$ and the centre of mass $x_i$ are calculated by integrat\-ing\- Eq.~(\ref{forces}) once and twice, respectively.

For simplicity $fluctuations$ in (\ref{forces}) are neglected and thus the forces are reduced to a driving and a repulsive term $F_i=F^{drv}_i + F^{rep}_i$. 
It is assumed that in the one-dimensional case a pedestrian only interacts with the man  directly in front. Consequently (\ref{frep}) reduces to 
\begin{equation}
F^{rep}_i =  F^{rep}_{i,i+1}(x_{i+1},x_i,v_i) 
\label{frep1}
\end{equation}

According to \cite{Helbing1995} we choose the driving term 
\begin{equation}
F^{drv}_{i}= m_i \frac{v^0_i-v_i}{\tau},
\label{FDRIV}
\end{equation}

where $v^0_i$ is the intended velocity of the $i^{th}$ pedestrian and $\tau$ is the acceleration time. For simplicity we set $v^0_i>0$, $x_{i+1}>x_{i}$ and $m_i=1$.
The repulsive force is modeled as:  

\vspace{-0.4cm}

 \begin{eqnarray}
 F^{rep}_i(t) = \left\{\begin{array}{r@{\quad:\quad}l}
             -  \frac{v_i(t)}{\tau} & \mbox{if cond.}^*\\
             0 & \mbox{otherwise}
             
\end{array}\right.		
\label{FREP}
\end{eqnarray}
$^*$cond.: $(dist_{i,i+1}(t) \leq \alpha \land  \frac{d(dist_{i,i+1})}{dt} < 0) \lor\, (dist_{i,i+1}(t) > \beta \land  \frac{d(dist_{i,i+1})}{dt} > 0).\\$

The quantities $\alpha$\, and $\beta$\, are chosen such that $\alpha < \beta$. To avoide the case where two pedestrians pass each other, we assume that the $i^{th}$ pedestrian stops once she/he is in contact with the pedestrian in front, i.e. $dist_{i,i+1}=0.$

\section{Motivation for Event-Driven Simulation }

An event is a single occurrence of a change in the forces acting on a pedestrian e.g. when the $i^{th}$ pedestrian comes 'to close' to the pedestrian in front, the force  $F_i$ changes at a special time - $t_{event}$ - from $F_i^{drv}$ to $F_i^{driv}+F_i^{rep}$.  The search for eventually occurring events is performed stepwise with a time bound $\Delta t$. It should not be so small as to make the simulation run too long, nor should it be so large as to make the number of events unmanageable. Since the forces acting on the pedestrians do not change between two events (event1 and event2), on can integrate Eq. (\ref{forces}) from $t_{event1}$ to $t_{event2}$ to calculate the velocity and the position of the pedestrians in the interval $[t_{event1} \,\,t_{event2}]$.

 %%%%%%%%%%%%%%%%%%%%%
 This approach uses a list of events that occur at various times, and handles them in order of increasing time. The simulation makes time 'jump' to the time of the next event. The simulation proceeds event-by-event rather than step-by-step.

\section{Event-driven simulation with velocity-adaptation}
 \label{section2}
The pedestrians shall adjust their actual velocity in such a way that they keep a minimum distance to the pedestrian in front. Again the forces are only influenced by actions in front of a pedestrian. At the moment where the distance between the $i^{th}$ pedestrian and the pedestrian in front reaches decreasingly $\alpha$, the $i^{th}$ pedestrian is subject to the repulsive force  (\ref{FREP}). This leeds to a reduction in the velocity $v_i$.
Once the distance between the $i^{th}$ pedestrian and the pedestrian in front reaches increasingly $\beta$ the action of (\ref{FREP}) vanishes. The velocity of the $i^{th}$ pedestrian increases again.

%The velocity function of a pedestrian is shown in Figure (\ref{fig:vel2}): 

 %\begin{figure}[thb]
  %\centerline{
%\includegraphics[width=0.485\columnwidth]{velocity2.eps}
%\hspace{0.3cm}
%    \qquad
 %\includegraphics[width=0.485\columnwidth]{man.eps}}
  % \label{velMan}
  %\caption{Fundamental diagrams for pedestrian %movement ....}
%\end{figure}

\begin{figure}[ht]
\centerline{\epsfig{file=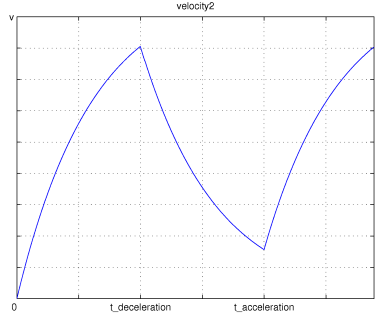,height=4.6cm,angle=0}}
\label{fig:vel2}
\caption{The variation of the velocity of a pedestrian}
\end{figure}

\begin{figure}[ht]
\centerline{\epsfig{file=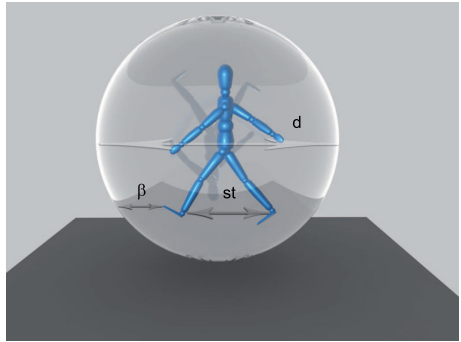,height=4.5cm,angle=0}}
\label{fig:man}
\caption{Illustration of $\alpha$, $\beta$ and $d$ }
\end{figure}

%%%####%%%
As the required length $d$ is modeled as a linear function of the velocity, the quantities $\alpha$ and $\beta$ are chosen to be linearly velocity-depended. $\beta$ is defined as the difference between the required length $d_i$ and the step-length $st_i(t)$ of the $i^{th}$ pedestrian and $\alpha$ as $\frac{\beta}{2}$. See Fig.~(\ref{fig:man}).

Because of
\begin{equation}
st_i = st_a + st_b\,v_i \quad \mbox{with} \quad st_a=0.235\,m \quad \mbox{and} \quad st_b=0.302\,s
\end{equation}

according to Ref. \cite{Weidmann1993} and
Eq. (\ref{dab}) one gets the complete definition of $\beta = \beta_a + \beta_b\,v$ ($\beta_a=0.125\,m,\;\beta_b=0.758\,s$) and thus of $\alpha=\alpha_a + \alpha_b\,v$ ($\alpha_a=0.0625\,m,\;\alpha_b=0.392\,s$). 
%as
%
%\begin{equation}
%\beta = \beta_a + \beta_b\,v \quad \mbox{with} \quad \beta_a=0.125\,m \quad \mbox{and} \quad \beta_b=0.758\,s
%\end{equation}
%
%and 
%
%\begin{equation}
%\alpha = \alpha_a + \alpha_b\,v \quad \mbox{with} \quad \alpha_a=0.0625\,m \quad \mbox{and} \quad \alpha_b=0.392\,s\,,
%\end{equation}

%respectively. 
 
%In Figure \ref{fig:man} the quantities $\beta$, $st$ and $d$ are illustrated.

From the definition of the repulsive force (\ref{FREP}) one recognizes that times, at which the distance between two pedestrians is $0$\,, $\alpha$ or $\beta$, are times where events occur, i.e. at those times force-variations happen. To calculate those events, one solves the equation: 

\begin{equation}
dist_{i}(t)= \zeta \, , \zeta \in \{ 0,\,\alpha\,, \beta\}
\label{cda}
\end{equation}

% For $\zeta=0$ the solution of equation (\ref{cda}) determines the collision time ($t_{collision}$): Time at which distance $dist_i(t)$ between the $i^{th}$ pedestrian  \,and the $(i+1)^{th}$ pedestrian is zero. For $\zeta=\alpha$ the deceleration time ($t_{deceleration}$): Time at which the distance $dist_i(t)$ between the $i^{th}$ pedestrian  \,and the $(i+1)^{th}$ pedestrian is $\alpha$ is calculated. And  for $\zeta=\beta$  the acceleration time ($t_{acceleration}$): Time at which the distance $dist_i(t)$ between the $i^{th}$ pedestrian  \,and the $(i+1)^{th}$ pedestrian is $\beta$ is calculated.

\section{Results}

To compare the model result with the empirical fundamental diagram of the single-file 
movement \cite{Seyfried2005}, a periodic passageway of length $L=17.3$\,m is selected. 
The values for the intended velocity $v^0_i$ is distributed according to a 
normal-distribution with a mean value of $\mu=1.24$\,m/s and $\sigma=0.05$\,m/s . According to Helbing \cite{Helbing1995}, $\tau=0.61$\,s is a reliable value. 

For every run at $t=0$ all velocities are set to zero and
the pedestrians are uniformly distributed. After $300$\,s 
relaxation, measurement are performed $300$\,s long. The determination of velocity mean value over all pedestrians is done after every event-evalutation. 

\begin{figure}[ht]
\centerline{\epsfig{file=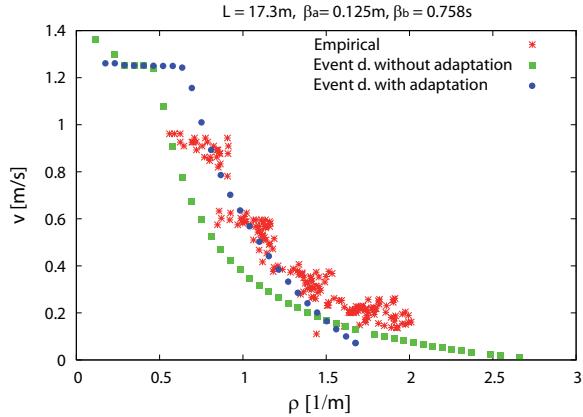,height=5.5cm,angle=0}}
\label{HDSPH}
\caption{Results: The fundamental diagram}
\end{figure}

Figure (\ref{HDSPH}) presents the dependency between mean velocity and density for different approaches to the velocity function (with and without adaptation). To demonstrate the influence of the velocity dependence of the required length different values for the parameter $b$ were selected. 

With $b=1.06$\,s and the repulsive force (\ref{FREP}) a good agreement at low densities between the velocity-density relation predicted by the model and the empirical 
fundamental diagram is obtained. However at high densities the velocity-density relation predicted by the model is inaccurate. 

\vspace{-0.3cm}

\section{Discussion and summary}

From the empirical fundamental diagram one determines $b=1.06$\,s and $a=0.36$\,m, see \cite{Seyfried2005}. The parameters for the step-length $st$ are also empirically known, see \cite{Weidmann1993}. Thus the only free parameter of the model is $\alpha$. The concept of velocity adaptation and the empirically known parameters lead to a good agreement with the fundamental diagram at low densities if one chooses $\alpha=\frac{\beta}{2}$

However at high densities one observe a discrepancy between the obtained results and the fundamental diagram. This can be explained by the fact, that at high densities the reaction-time and the behavior of pedestrians become more random so that the deterministic character of the model don't reflect anymore the stochastic behavior of the pedestrians. An improvement of the model should include a stochastic in the reaction-time of pedestrians and determining an appropriate ratio  of  $\alpha$\, to\, $\beta$ (here it was set to $\frac{1}{2}$).

%\bibliography{ped}

\end{document}